\documentclass[a4paper,11pt]{article}
\usepackage{pos}

\title{Supernova Remnants in the IXPE era: a review}
\ShortTitle{Supernova Remnants in the IXPE era: a review}

\author*[a]{Riccardo Ferrazzoli}
\author[b]{the IXPE Collaboration}

\affiliation[a]{INAF Istituto di Astrofisica e Planetologia Spaziali,\\
Via del Fosso del Cavaliere 100, 00133 Roma, Italy}

\affiliation[b]{https://ixpe.msfc.nasa.gov/partners$\_$sci$\_$team.html}

\emailAdd{riccardo.ferrazzoli@inaf.it}

\abstract{The Imaging X-ray Polarimetry Explorer (IXPE) has opened a new observational window on the physics of supernova remnants (SNRs) by providing the first spatially resolved X-ray polarimetry measurements. 
These data directly probe the geometry and turbulence of magnetic fields in regions of efficient particle acceleration, thereby constraining models of diffusive shock acceleration and magnetic-field amplification. 
IXPE has so far observed six young SNRs  (Cas A, Tycho, SN 1006, RX J1713.7$-$3946, Vela Jr., and RCW 86) with published results on the first five. 
The observations reveal significant polarization in all cases, with degrees of polarization ranging from $\sim5$\% to over 30\%, reflecting different turbulence levels and environmental conditions. 
Three remnants (Cas A, Tycho, and SN 1006) show predominantly radial magnetic fields, while RX J1713.7-3946 and Vela Jr. display tangential morphologies. 
This dual behavior, not simply correlated with evolutionary stage, challenges the long-standing dichotomy inferred from radio observations and suggests that both shock velocity and circumstellar medium density play key roles in shaping magnetic-field topology. 
IXPE’s results mark a major step toward disentangling the processes governing cosmic-ray acceleration in young SNR shocks, with ongoing and future observations expected to further constrain the interplay between turbulence, shock dynamics, and particle acceleration.
}

\FullConference{Frontier Research in Astrophysics – IV (FRAPWS2024)\\
 9-14 September 2024\\
Mondello, Palermo, Italy\\}

\begin{document}
\maketitle

\section{Introduction}
The NASA-ASI Imaging X-ray Polarimetry Explorer (IXPE) \cite{2022Weisskopf, 2021Soffitta} is the first mission entirely dedicated to spatially resolved X-ray polarimetry, with a spatial resolution of $\sim30"$ in the 2 - 8 keV energy band.
In the last four years, IXPE has been doing a unique science, observing and studying extended sources such as supernova remnants (SNRs).
SNRs are the result of the violent death of a star whose ejected matter interacts with the interstellar medium and produces shocks. 
One of the major points of interest for the study of these objects is that their shocks are thought to be the dominant source of Galactic Cosmic Rays up to energies of $\geq100$ TeV.
In these shocks, particles are accelerated through the mechanism of diffusive shock acceleration (DSA, e.g. \cite{2001Malkov}).
DSA results in a non-thermal population of high-energy particles whose maximum energy being is limited by radiative losses, the SNR age, or particle escape \cite{2022Vink_review}.
In this process, particles are scattered back and forth across the shock, gaining energy each time they cross the shock.
For this process to be efficient, strong and turbulent magnetic fields are needed, with the turbulence being either preexisting or self-generated, coming from the streaming ions themselves.
The mean free path of the particles is a multiple of its gyroradius and the so called Bhom factor $\eta$ that is connected to how efficient the acceleration process is.
For the loss-limited case, the maximum acceleration efficiency is achieved for $\eta=1$, for which the mean free path is the smallest.
The maximum energy that one can get from these particle accelerators is proportional to the shock velocity and inversely to the magnetic field, so that for typical values for young SNRs one gets tens of TeV.
Turning the maximum energy into a synchrotron spectrum, the associated cut-off falls at an energy that is going to be again related to the shock velocity and the Bohm factor meaning that if one can measure a spectrum and look at the cut-off, knowing the shock velocity, it is actually possible to measure the Bohm factor and figure out how efficient the particle acceleration is.

\begin{figure}[htbp]
	\centering
	\includegraphics[width=1.0\linewidth]{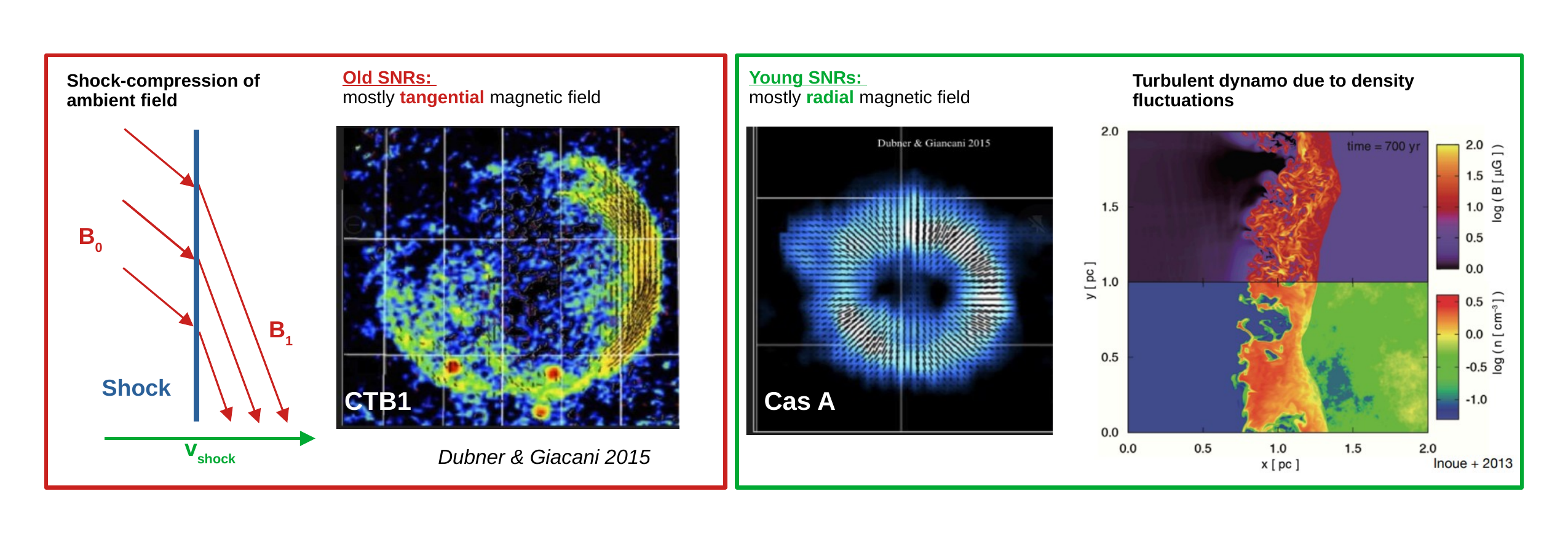}
 \caption{Schematic representation of the magnetic field dichotomy observed in the radio energy band between old (tangential) and young (radial) SNRs.
 \textbf{(Left):} example of old SNR with tangential magnetic field in the radio band and scheme of tangential shock compression of the upstream ambient magnetic field. 
 \textbf{(right):} example of young SNR with radial magnetic field in the radio band and scheme radial turbulent dynamo due to density fluctuations 
 Images adapted from \cite{2015Dubner, 2013Inoue}.}
 \label{fig:dichotomy}
\end{figure}

High resolution Chandra images found in most young SNRs shocks that the X-ray emission comes from thin, $\sim$few arcseconds, filaments that we ascribe to synchrotron emission.
This means that a population of relativistic electrons must exist that is accelerated very close to the shocks, on spatial scales of $\sim10^{17}$cm.
Synchrotron radiation is intrinsically polarized, so even if IXPE, with its 30 arcseconds angular resolution, cannot resolve these filaments like Chandra does, we know that the X-rays and hence the polarized emission comes from close to the shocks.
This is in contrast with, for example, radio-emitting electrons that instead traveled far from the acceleration regions.
X-ray polarimetry allows us to determine the level of order of the magnetic fields and their orientation by measuring the degree of polarization and the polarization angle - that is orthogonal to its direction - respectively. \\
Indeed, the morphology of the magnetic field of SNRs is something we are very interested in because of a dichotomy that we observe in the radio band: old remnants tend to have tangential magnetic fields, whereas young SNR such as Cas A, or Tycho have radial magnetic fields (see Fig. \ref{fig:dichotomy}).
The former morphology can be easily interpreted as compression of the magnetic field component that is orthogonal to the shock motion as it sweeps the upstream turbulence.
An explanation for the latter is less clear.
One hypothesis argues for hydro-dynamical instabilities such as Rayleigh-Taylor filaments that are strung along and stretch out so that the magnetic field is carried along in those directions and the particles are spiraling along that magnetic field and that is what gives the radial orientation \cite{2013Inoue}.
A second one calls for a sort of selection effect due to a more efficient particle acceleration in regions where the shock velocity is parallel to the magnetic field \cite{2017West}.
So the question is: why does this happen, and is it the same at smaller scales where X-rays are emitted?

As for the PD, and hence the turbulence, before the launch of IXPE predictions were attempted for the observational expectations from different characteristics of the turbulence \cite{2009Bykov, 2020Bykov}.
For a steep turbulence spectrum, meaning that there is higher amplitude turbulence at small spatial scales, it would be possible to detect polarization propagating downstream.
If the spectrum is relatively flat then the polarization would be harder to see.
On the other hand, if there are built-in anisotropies in the turbulence spectrum as it is crossing the shock, depending on the scale of these anisotropies, large polarization degrees might be resolved with IXPE. \\
As of now, IXPE observed six young SNRs: Cas A, Tycho, SN1006, RX J1713, Vela Jr, and RCW 86 reporting results for all but the latter.
In te following we summarize the major results for each one of them.

\subsection{Cas A}
\begin{figure}[htbp]
	\centering
	\includegraphics[width=1.0\linewidth]{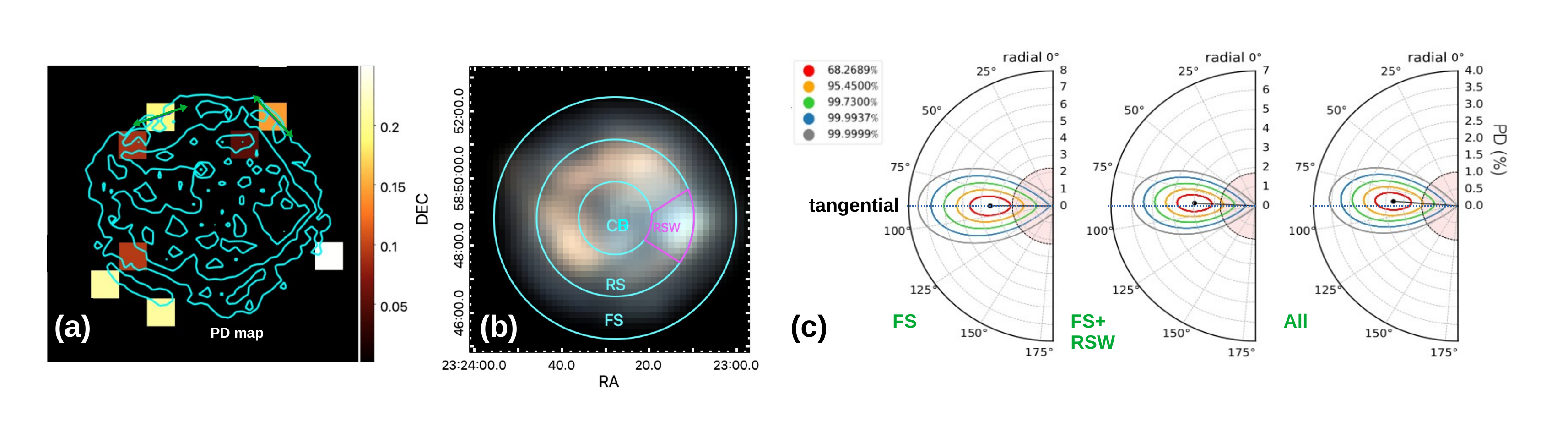}
 \caption{\textbf{(a):} Polarization map in the 3 - 6 keV energy band binned on a 42" pixel size of the SNR Cas A. Only pixels with confidence levels above $2\sigma$ are shown. For pixels with $3\sigma$ confidence level) the polarization angles are indicated with green arrows.
 \textbf{(b):} IXPE three color Stokes I image with superimposed the regions of interest used to test for an overall radial or tangential magnetic field orientation. The colder colors indicates a stronger non-thermal emission with respect to the warmer ones.
 \textbf{(c):} Polar diagrams of the measured PD and PA with respect to circular symmetry as confidence contours for the regions of panel \textbf{b}. The shaded pink region corresponds to the MDP99 level. Values around 90 degrees correspond to a tangentially oriented PA averaged over the region, while around 0◦ indicates on average a radially oriented PA.
 Images adapted from \cite{2022Vink_b}.}
 \label{fig:CasA}
\end{figure}

Cas A was the first target of the IXPE scientific campaign after the launch and commissioning.
It is a bright 350 years old remnant resulting from a core collapse explosion with its spectrum dominated by bright emission lines and by a non-thermal continuum \cite{1995Reed,1998Vink}.
It also exhibits a clear reverse shock. 
The IXPE results were presented in \cite{2022Vink_b}.
A pixel-by-pixel search of the polarization signal, considering spatial bins of size 42'' and 84'' (see Fig. \ref{fig:CasA} (a)) revealed weak evidence for X-ray polarization of $\rm PD =5 - 15$\%, and tangential PA, suggesting radial magnetic fields. 
In order to improve the sensitivity, circular symmetry of the polarization direction can be assumed, and exploiting the additivity of the Stokes parameters, larger emission regions could be combined.
This method established highly significant detections of polarization from multiple regions: the forward shock, the forward shock plus a distinct region on the western edge of the reverse shock, and the whole remnant.
This is illustrated in Fig. \ref{fig:CasA} (b) and (c) showing a three-color IXPE image of Cas A with these regions identified and their polarization plots of the most significative results, respectively.
In the polarization plots, the PD is represented by the radial component of the polar plot, and the PA is measured relative to the radial direction. 
The contours show confidence intervals, and the pink circle illustrates the minimum detectable polarization at 99\% confidence. 
The PA values for the forward shock (FS), the forward shock plus the western part of the reverse shock (FS+RSW), and for the whole remnant (All) are consistent with tangential polarization -- and thus with radial magnetic fields -- with PD values ranging from $\sim 2-4$\%. 
Correcting for the dilution due to the thermal flux, determined using Chandra spectral modeling of the regions of interest, the PD values of the X-ray synchrotron emission alone are no higher than $\sim4\%$, that is, the value measured in the radio band.
The low PD measured is suggestive of high levels of turbulence in the region close to the shock.
The measured PA instead implies that that whatever the process that is responsible for the radial magnetic field observed in the radio, is already at work very close to the shock where particle are accelerated, on spatial scales corresponding to the thin rims ($\sim 10^{17}$ cm).

\subsection{Tycho}
\begin{figure}[htbp]
	\centering
	\includegraphics[width=1.0\linewidth]{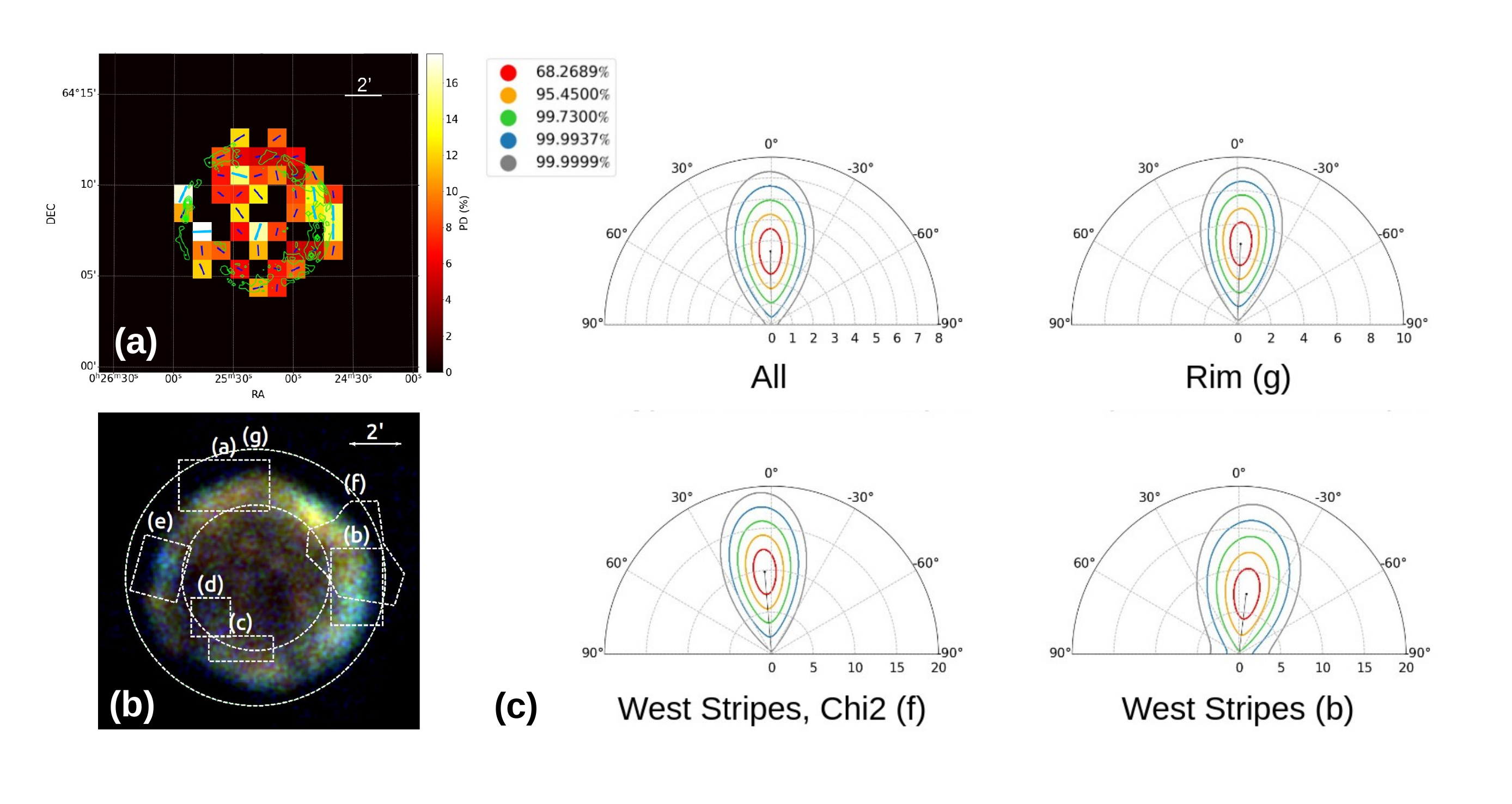}
 \caption{\textbf{(a):} Polarization map in the 3–6 keV energy band with a 1 arcminute pixel size. Only the pixels with significance higher than $1\sigma$ are shown. The blue bars represent the PA. The thicker
cyan bars mark the pixels with significance higher than $2\sigma$. Superimposed in green are the 4–6 keV Chandra contours.
 \textbf{(b):} IXPE three-color image of Tycho, superimposed are the regions of interest.
 \textbf{(c):} Polar plots for the most significative Tycho regions of interest. Values more consistent with a direction of polarization of 0 degrees correspond to an overall tangentially oriented polarization averaged over the region of interest. 
 Images adapted from \cite{2023Ferrazzoli}.}
 \label{fig:Tycho}
\end{figure}

The Tycho SNR is the result of a Ia explosion, observed as the historical supernova SN 1572 \cite{2003Green}.
A fascinating peculiarity of Tycho are striking stripe-like synchrotron structures, highlighted by Chandra observations \cite{2011Eriksen}, whose origin is not yet well understood but theoretical studies suggested that they can be highly polarized  \cite{2009Bykov,2011Bykov_a,2020Bykov}. 
Tycho was the second SNR observed by IXPE \cite{2023Ferrazzoli}. 
The analysis followed a similar approach to that used for Cas A, beginning with a pixel-by-pixel search for a signal.
However, since Tycho is not as bright as Cas A, the polarization map binned on a 1 arcminute scale does not show highly significant detections (see Fig. \ref{fig:Tycho} (a)).
By aligning and summing over the data from different regions of interest, such as the highest significant region in the west, the rim, but also the whole remnant as shown in of Fig. \ref{fig:Tycho} (b), highly significant detections of polarized emission were identified.
The measured tangential PA corresponds to a radial magnetic ﬁeld, consistent with the radio band polarization observations but originating from regions even closer to the shock.
After accounting for the dilution by the thermal component, the X-ray PD in Tycho is significantly higher than for Cas A, $12\%\pm2\%$ in the rim and $9\%\pm2\%$ in the whole remnant (see the polar plots in Fig. \ref{fig:Tycho} (c) for the measured values in selected regions), suggesting maybe a lower turbulence level or longer turbulence scales.

\subsection{SN 1006}
\begin{figure}[htbp]
	\centering
	\includegraphics[width=1.0\linewidth]{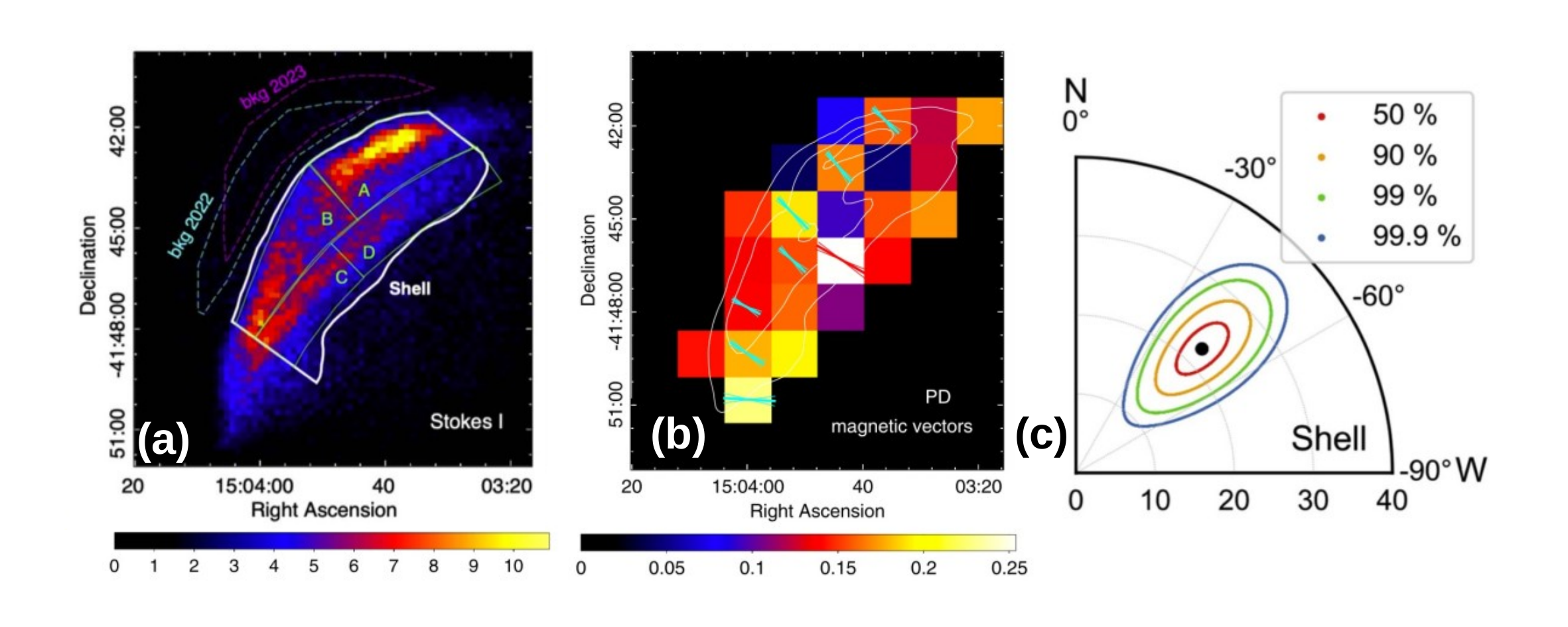}
 \caption{\textbf{(a):} IXPE Stokes I image in the 2–4 keV band of the SNR SN 1006 NW, with the regions of interest highlighted. The regions delineated with dashed lines correspond to the background regions. 
 \textbf{(b):} PD map overlaid with magnetic field vectors and their $1\sigma$ uncertainties. 
 The blue and red vectors correspond to pixels with significance $>2\sigma$ and $>3\sigma$, respectively.
 \textbf{(c):} polar plot of the Shell region after background subtraction.
 Images adapted from \cite{2023Zhou}.}
 \label{fig:SN1006}
\end{figure}

\begin{figure}[htbp]
	\centering
	\includegraphics[width=1.0\linewidth]{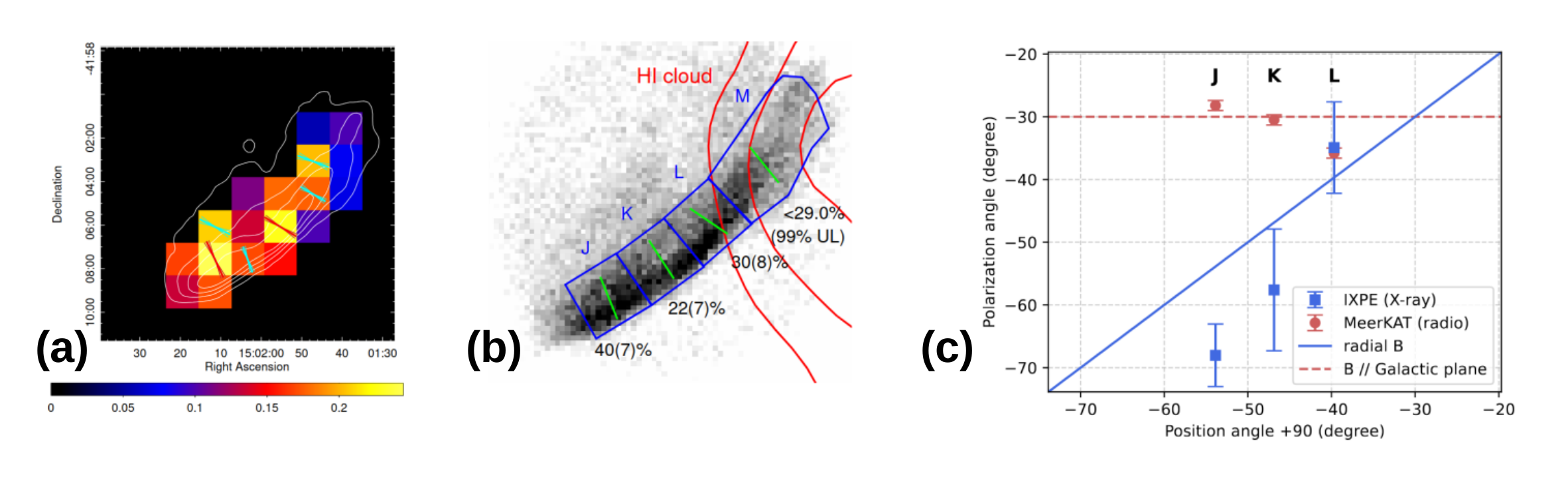}
 \caption{\textbf{(a):} IXPE PD map of the SW region of SN 1006 overlaid with magnetic field vectors and their $1\sigma$ uncertainties.
 \textbf{(b):} Regions selected in the polarization analysis on top of Stokes I image. The average magnetic field directions are shown with green lines. The flux between 5.7-10.7 km s$^{-1}$ of the HI cloud is marked by the red contours.
 \textbf{(c):} PA measured in 3 regions od SN 1006 SW with IXPE and MeerKAT compared to the position angles of the regions. The solid and dashed lines represent two magnetic configurations: radial magnetic orientation for the X-rays andd magnetic fields parallel to the Galactic plane in the radio. The error bars show the $1\sigma$ uncertainty.
 Images adapted from \cite{2025Zhou}.}
 \label{fig:SN1006_SW}
\end{figure}

The third remnant observed by IXPE was SN1006 \cite{2023Zhou}.
Differently for the others, this one has a large angular extension, and indeed the IXPE field of view can cover only the northeastern limb of SN 1006.
However, its spectrum is completely non-thermal \cite{1995Koyama}, making it an ideal target for X-ray polarimetry.
Compared to Cas A and Tycho, SN 1006 is significantly more extended, but fainter in the X-rays, hence background contributions from the source region were substantial, requiring their removal during analysis.
IXPE resolved the double-rim structure of SN 1006 NE, enabling a detailed spatial analysis of the X-ray polarization. 
The Stokes I image of SN 1006 NE in the 2 – 4 keV range is shown in Fig. \ref{fig:SN1006} (a), with the white and green regions indicating the shell and four sub-scale areas, respectively. 
The X-ray polarization of the entire shell was detected with a significance of $6.3\sigma$, with $\rm PD = 22.4\pm 3.5\%$ and $\rm PA = -45.5^\circ$, indicative of a radial magnetic field also in this case, and an even lower turbulence than the previous remnants. 
The sub-regions showed no significant variations of the polarization properties among them.
IXPE observed again SN1006 in 2024, this time focusing on the south-western limb \cite{2025Zhou}.
SN 1006 resides above the Galactic plane in a rarefied environment, but the southwestern limb is interacting with a hydrogen cloud (see Fig. \ref{fig:SN1006_SW}). 
In this limb, the IXPE observations reveal a robust detection of X-ray polarization across the 2–4 keV band, with an average polarization degree of $21.6 \pm 4.5\%$.
This global measurement closely matches the NE limb values (PD $\approx 22.4\%$), so on large scales the magnetic field orientations around the remnant’s shell is uniform.
However, with a more detailed spatial analysis, IXPE found that in the region J, located toward the southern edge of the SW limb, the PD peaks at $40 \pm 7\%$ ($5.7\sigma$), indicating an exceptionally ordered magnetic field immediately behind the shock front. 
Conversely, in sector M — coincident with the region where the shock encounters a dense HI cloud — the polarization is markedly diminished, with a 99\% upper limit of PD$<27\%$. 
This stark contrast suggests that denser ambient material enhances magnetic turbulence and disorder in the post-shock region, thereby reducing the observed polarization.
Comparing the X-ray results to MeerKAT radio polarimetry at L-band reveals a significant offset: the radio PA of $\approx–29°$ aligns with the Galactic plane, whereas the X-ray PA is more radial. 
This discrepancy highlights the different scales probed by each wavelength regime—X-rays trace the freshly accelerated electrons interacting with small-scale, post-shock magnetic fluctuations, while radio emission arises from older electrons diffusing through the larger-scale, ambient interstellar field.

\subsection{RX J1713.7-3946}
\begin{figure}[htbp]
	\centering
	\includegraphics[width=1.0\linewidth]{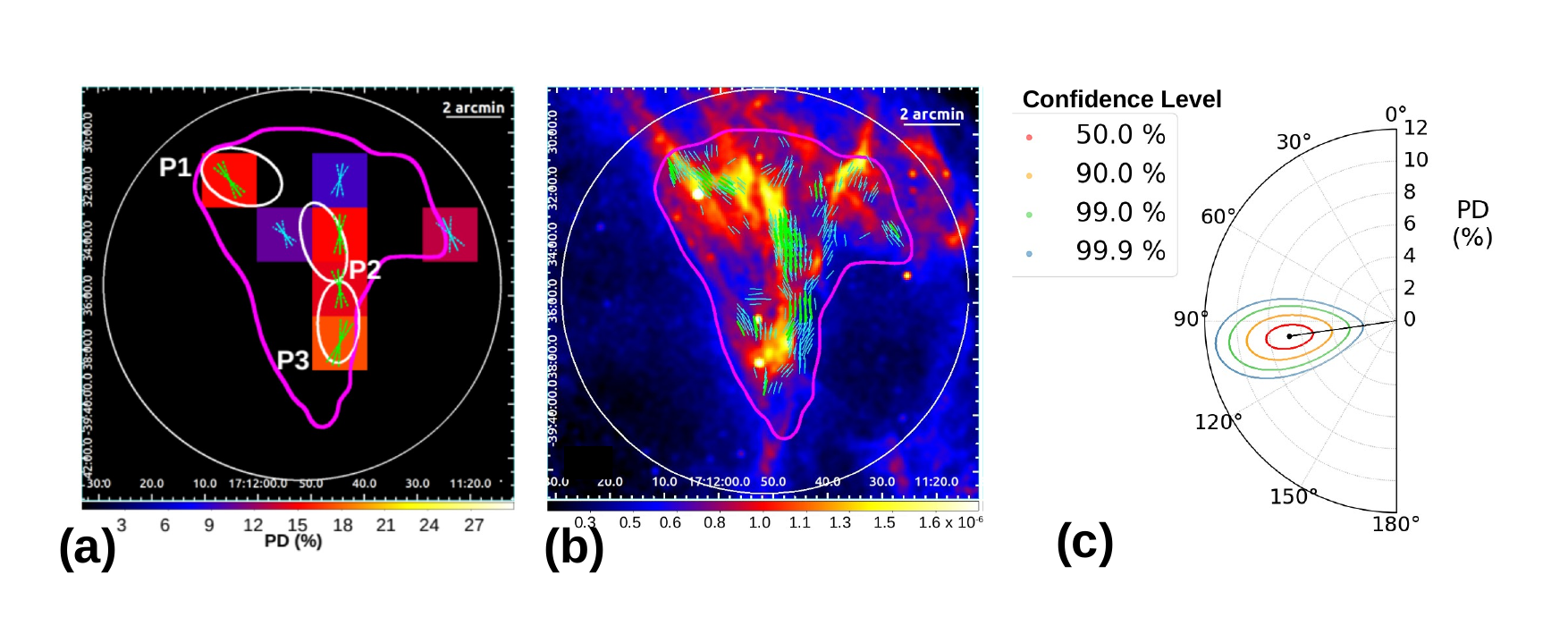}
 \caption{\textbf{(a):} polarization map binned with 2' wide pixels. 
 The cyan and green vectors represent the direction of the magnetic field revealed at 2$\sigma$ and 3$\sigma$ significance levels, respectively. 
 The length of the vectors is proportional to the PD. 
 The dashed vectors show the 2$\sigma$ uncertainty on the magnetic field direction in each pixel to improve the visibility. 
 \textbf{(b):} Chandra exposure-corrected mosaic of 0.5-7 keV images of the northwest of RX J1713 with the IXPE field of view as a white circle, the IXPE 2-5 keV contours in magenta. 
 In cyan and green, we show the magnetic field lines obtained through Gaussian smoothing of the IXPE data with >2$\sigma$ and >3$\sigma$, respectively.
 \textbf{(c)}: Polarization plot of the whole NW shell in the 2-5 keV energy band.
 Figures adapted from \cite{2024Ferrazzoli}.
 }
 \label{fig:RXJ1713}
\end{figure}

RX J1713.7-3946 is a large ($\sim1$ degree in diameter), entirely non-thermal \cite{1997Koyama,1999Slane} shell-type SNR located in the Galactic plane. 
It is believed to have resulted from a Type Ib/c supernova, potentially linked to the historical SN 393 event \cite{1997Wang,2016Tsuji,2017Acero}, making it the oldest SNR whose results have been reported by IXPE to date \cite{2024Ferrazzoli}. 
IXPE observed the northwestern portion of this remnant.
In Fig. \ref{fig:RXJ1713} the binned and smoothed polarization map, the latter overlapped to the high-resolution Chandra image, and the polarization result for the whole NW shell are shown.
Analysis of both the polarization map and regions of interest revealed that the polarization direction is perpendicular to the shock, with an average polarization degree of $12.5\pm3.3\%$ across the entire region. 
Unlike other IXPE-observed remnants, which exhibit radial magnetic fields, RX J1713.7-3946 is the first case where shock-compressed tangential magnetic fields dominate in the X-ray band. These results align with a model where shock compression of upstream isotropic turbulence generates a primarily tangential magnetic field.

\subsection{Vela Jr.}
Vela Jr. is another large and non-thermal remnant residing in a corner of the even larger Vela SNR.
This remnant has a Core collapse origin, and it is even older than RX J1713, with an estimated age between 1700 and 4300 years, and again has a non thermal spectrum.
Also in this case, similarly to RX J1713, IXPE found a significant detection of an overall tangential magnetic field, with an average PD $= 16.4 \pm 5.2\%$ \cite{2024Prokhorov}.
This is also the first time the magnetic field of this object was mapped at any frequency, and again it challenges the simple dichotomy that young remnant have radial field, and old one tangential, as both Vela Jr. and RX J1713 are dynamically young.
\begin{figure}[htbp]
	\centering
	\includegraphics[width=1.0\linewidth]{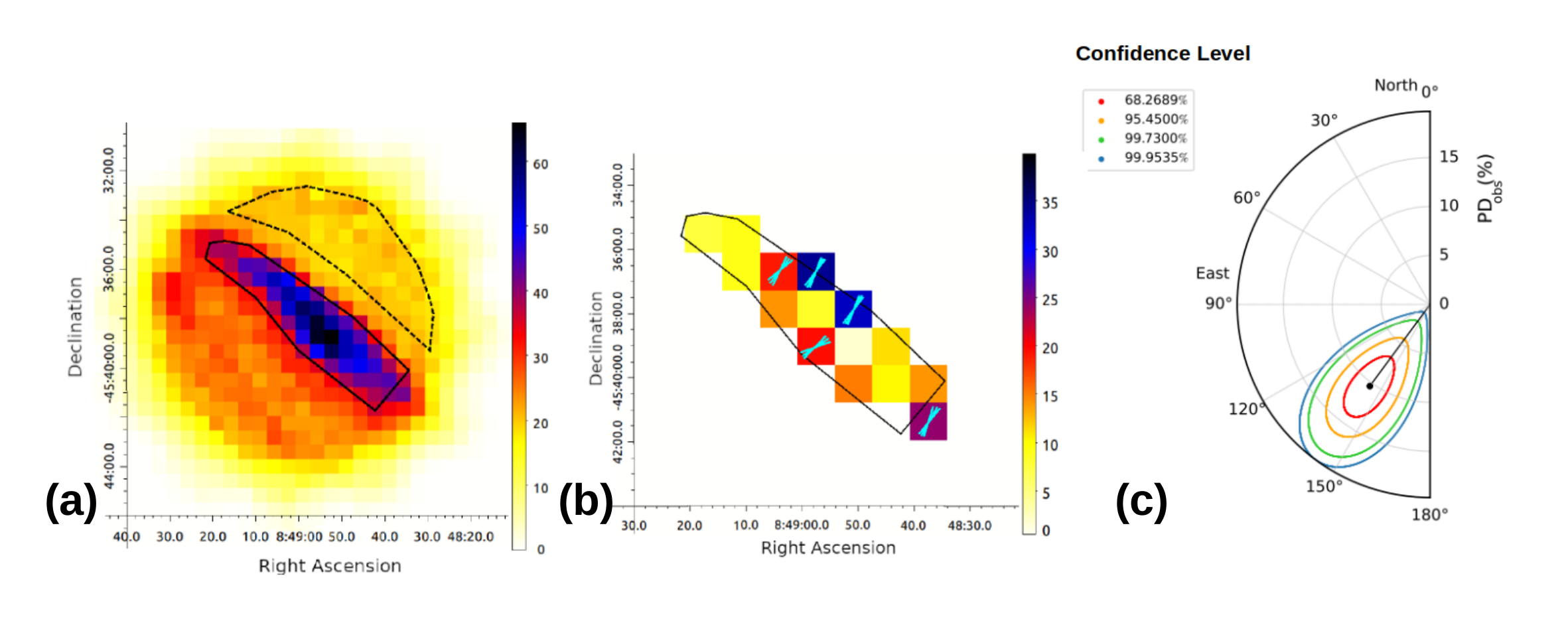}
 \caption{\textbf{(a):} IXPE Stokes I map of Vela Jr. with a pixel size of 30''. The map shows the source and background regions by solid and dashed lines, respectively.
 \textbf{(b):} Polarization map overlaid with polarization vectors and their $1\sigma$ errors on PA. The vectors correspond to pixels with a pretrial significance above $2\sigma$.
 \textbf{(c):} Polar plot obtained from the analysis of the source region.
 Images adapted from \cite{2024Prokhorov}.}
 \label{fig:VelaJr}
\end{figure}

\subsection{Current status and conclusions}
According to the current picture of the X-ray polarimetry of SNR we have significant detection of polarization in four sources: Cas A, Tycho, SN1006, and RX J1713.7-3946.
The former three have radial magnetic field, while the latter has a tangential one.
The PD of the synchrotron emission ranges between 5\% – 30\%, pointing to high level of turbulence, as expected for efficient particle acceleration by DSA.
The values vary between the SNRs, and there may be trends with the ambient density playing a role, or the Bhom factor (see table \ref{tab:snr_characteristics}). 
\begin{table}[htbp] \small
    \centering
    \begin{tabular}{lcccccccc}
        \hline
        SNR & \multicolumn{3}{c}{Polarization Degree (\%)} & $v_{\text{shock}}$ & $n_0$ & $\eta$ & $B_{\text{low}}$ &   \\
        & Rim & SNR & Peak & (km s$^{-1}$) & (cm$^{-3}$) & & $(\mu \text{G})$ & orient.  \\
        \hline
        Cas A        & 4.5  $\pm$ 1.0 & 2.5 $\pm$ 0.5 & $\sim$15    & $\sim$5800 & 0.99 $\pm$ 0.3  & $\sim$1--6 & 25--40  & R \\
        Tycho        & 12   $\pm$ 2   & 9 $\pm$ 2     & 23 $\pm$ 4  & $\sim$4600 & 0.1--0.2        & 1.5--2.5   & 30--40  & R \\
        SN 1006 (NE) & 22.4 $\pm$ 3.5 & \dots         & 31 $\pm$ 8  & $\sim$5000 & 0.05--0.08      & 6--18      & 18--26  & R \\
        SN 1006 (SW) & 21.6 $\pm$ 4.5 & \dots         & \dots       & $\sim$5000 & 0.05--0.16      & 9--20      & 15--30  & R \\
        RX J1713 (W) & 13.0 $\pm$ 3.5 & \dots         & 36 $\pm$ 10 & 1400--2900 & $\sim$0.01--0.2 & 1.4--8.6   & 30--100 & T \\
        Vela Jr (NW) & 16.4 $\pm$ 5.2 & \dots         & 55--78      & $\sim$3000 & $\leq$0.01      & $\sim$1    & 30--80  & T \\
        \hline
    \end{tabular}
    \caption{X-ray polarization and shock properties for young SNRs. R means radial magnetic field and T means tangential magnetic field, n0 is the upstream ambient density. $\eta$ is the Bhom factor and B$_{low}$ is the estimate of the downstream magnetic field.}
    \label{tab:snr_characteristics}
\end{table}
It is noteworthy that the level of circumstellar medium interaction - indicated by the ambient density n0 - is high in Cas A, modest in Tycho, and low in SN 1006, RX J1713 and Vela Jr.
The PD varies in the opposite sense, decreasing with increasing CSM interaction. 
It is conceivable that high levels of turbulence are associated with higher CSM densities, and that this results in a low polarization degree downstream.
Must importantly, however, we have two families of magnetic field morphology: radial in Cas A, Tycho, and SN 1006; tangential for RX J1713 and Vela Jr.
And this is more puzzling, as these are all dynamically young remnants, but behaving like the old and young remnants in the radio band.
A recent study \cite{2024Bykov} suggested an explanation at least for the different morphology observed in the magnetic fields based on magneto-hydrodynamical simulations.
The idea is that the cosmic ray current generates magnetic and density fluctuations through Bell instability. 
Upon passing through the shock, these density fluctuations generate anisotropic vortex turbulence that produces a radial magnetic field component.
The simulations show that the observed magnetic field morphology depends on where this radial magnetic field component peaks: for high shock velocities, it peaks close to the shock and ends up being the dominating component.
This is the case for Cas A, Tycho, and SN 1006.
On the other hand, lower shock velocities cause the radial component to peak farther downstream, at distances larger than the synchrotron cooling distance, and we sample just the shock-compressed tangential component near the shock.
And this would be the case of RX J1713 and Vela Jr.

IXPE studies of additional SNRs, with a range of ages and inferred Bohm factors, are underway to provide additional constraints on the connection between polarization properties and the remnant conditions. 
Measurements of the polarization degree and magnetic field geometry will continue to provide new insights into the conditions leading to efficient particle acceleration in fast shocks.

\section*{Acknowledgments}
The Imaging X-ray Polarimetry Explorer (IXPE) is a joint US and Italian mission. 
The US contribution is supported by the National Aeronautics and Space Administration (NASA) and led and managed by its Marshall Space Flight Center (MSFC), with industry partner Ball Aerospace (contract NNM15AA18C). 
The Italian contribution is supported by the Italian Space Agency (Agenzia Spaziale Italiana, ASI) through contract ASI-OHBI-2022-13-I.0, agreements ASI-INAF-2022-19-HH.0 and ASI-INFN-2017.13-H0, and its Space Science Data Center (SSDC) with agreements ASI-INAF-2022-14-HH.0 and ASI-INFN 2021-43-HH.0, and by the Istituto Nazionale di Astrofisica (INAF) and the Istituto Nazionale di Fisica Nucleare (INFN) in Italy. 
This research used data products provided by the IXPE Team (MSFC, SSDC, INAF, and INFN) and distributed with additional software tools by the High-Energy Astrophysics Science Archive Research Center (HEASARC) at NASA Goddard Space Flight Center (GSFC).
R.F. is partially supported by MAECI with grant CN24GR08 “GRBAXP: Guangxi-Rome Bilateral Agreement for X-ray Polarimetry in Astrophysics”.

%
%
%
%
%
%
\end{document}